\documentclass[12pt]{article}

\usepackage{scicite}

\usepackage{times}

\usepackage{deluxetable}

\usepackage{graphicx}

\newenvironment{sciabstract}{%
\begin{quote} \bf}
{\end{quote}}

\newcounter{lastnote}

\title{The Shortest Known Period Star Orbiting our Galaxy's Supermassive Black Hole}

\author
{L. Meyer,$^{1}$ A. M. Ghez,$^{1\ast}$ R. Sch\"odel,$^{2}$ S. Yelda,$^{1}$ \\ A. Boehle,$^{1}$  J. R. Lu,$^{3}$ T. Do,$^{4, 5}$ M. R. Morris,$^{1}$ \\ E. E. Becklin,$^{1}$ K. Matthews$^{6}$\\
\\
\normalsize{$^{1}$University of California Los Angeles, Department of Physics and Astronomy,}\\
\normalsize{Los Angeles, CA 90095-1547, USA}\\
\normalsize{$^{2}$Instituto de Astrof\'isica de Andaluc\'ia (CSIC), Glorieta de la Astronomia S/N, 18008 Granada, Spain}\\
\normalsize{$^{3}$University of Hawaii, Institute for Astronomy,  Honolulu, HI 96822, USA}\\
\normalsize{$^{4}$University of California Irvine, Department of Physics and Astronomy,}\\
\normalsize{Irvine, CA 92697-4575, USA}\\
\normalsize{$^{5}$Dunlap Institute for Astronomy and Astrophysics, University of Toronto,}\\
\normalsize{Toronto M5S 3H4, Ontario, Canada; Dunlap Fellow}\\
\normalsize{$^6$California Institute of Technology, Division of Mathematics,}\\
\normalsize{Physics and Astronomy, Pasadena, CA 91125, USA}
\\
\normalsize{$^\ast$To whom correspondence should be addressed; E-mail:  ghez@astro.ucla.edu}\\
}

\date{}

\begin{document} 


\baselineskip24pt


\maketitle


\begin{sciabstract}

Stars with short orbital periods at the center of our galaxy offer a powerful and unique probe of
a supermassive black hole.  Over the past 17 years, 
the W. M. Keck Observatory has been used to image the Galactic center at 
the highest angular resolution possible today.  
By adding to this data set and advancing methodologies, we have detected S0-102, a star orbiting our galaxy's supermassive black hole with a period of just 11.5 years. S0-102 doubles the number of stars with full phase coverage and periods less than 20 years. It thereby provides the opportunity with future measurements to resolve degeneracies in the parameters describing the central gravitational potential and to test Einstein's theory of General Relativity in an unexplored regime.

\end{sciabstract}

%% Bring in  redshift

The recent advent of high resolution imaging capabilities on large ground-based telescopes has afforded us a  view of the center of our galaxy that is substantially altering our understanding of this region and by analogy the centers of other galaxies. 
The earliest high-resolution images in the near-infrared (NIR) revealed the presence of a  black hole with a mass $4\cdot 10^6$ times that of the sun
by detecting stars at the very heart of the galaxy, the so called S-stars, with velocities of up to $\sim$10,000~km/s ({\it 1, 2}). The detection of accelerations in the motion of these stars ({\it 3, 4}) further strengthened the case for a non-stellar dark mass. After passing periapse -- the closest point to the black hole in the orbit of a star --  a Keplerian orbit could be determined for the star S0-2 (with an orbital period of 16 years), proving the black hole hypothesis beyond reasonable doubt ({\it 5, 6}).
This star can now be traced in its motion around the black hole with such precision and complete phase coverage that its three-dimensional orbit gives the most accurate description to date of the central mass and distance from our Solar system to the center of the Galaxy ({\it 7--9}); no other star has so far been reported with more than $\sim$40\% of its orbit covered by data. 

Here, we report the detection of another star, S0-102, in orbit around the supermassive black hole at the center of our galaxy ({\it 10}). It has an orbital period of 11.5 years, which makes it the shortest period star known yet.

The data sets that we used for this study come from 
high-resolution imaging observations of the Galactic center
that were obtained at the W. M. Keck Observatory between 1995 and 2012 with speckle imaging and adaptive optics.
Speckle imaging with the near-infrared imaging camera NIRC ({\it 11, 12}) on Keck I provided the earliest high angular resolution imaging data sets (1995 - 2005), which are reported 
with their observational setup in 
several earlier papers ({\it 2, 3, 7, 13, 14}).  In summary, several thousand data frames
capture images of the Galactic center on timescales of $\sim$0.1 sec, which is short enough compared to the 
coherence time of atmospheric turbulence to freeze the distorting effects of the Earth's atmosphere and preserve the source's information at all spatial frequencies transmitted by the telescope. Table~S1 summarizes key observational information for the 26 epochs of speckle imaging data that are available.

Adaptive optics (AO) observations of the Galactic center were carried out between 2004 and 2012, with substantially more data per epoch being obtained in 2006 and onwards.  Previous publications contain the
observational details for the first 9 epochs of AO observations ({\it 7, 14--16}).  We collected 12 further epochs of AO observations between
2008 May and 2012 May, using an identical observational set-up to all AO observations obtained in 2006 and
onwards.  For all but two epochs, the high-order AO compensation was based
on observations of a bright (R$\sim$10 mag), on-axis laser guide star (LGS), generated with the Keck II sodium 
laser, and the low-order (e.g. tip-tilt) AO corrections were determined from measurements of the natural guide star (NGS) USNO 0600-28577051 (R = 13.7 mag and $\Delta r_{SgrA*}$ = 19$\tt''$). Table~S2 provides the details of the 20 AO epochs that are used here.

The data analysis for this study falls into several distinct stages ({\it 17}).
First, for each epoch of observations, we constructed an average image of all the high
quality data collected as well as three sub-set images, each composed of 
one third of the data (see Fig.~1 for a zoom into the region of an AO image where S0-102, S0-2, and the electromagnetic manifestation of the black hole, Sgr~A*, reside). Second, both previously known and newly detected stars were 
identified in the images, along with
their photometric and astrometric characteristics, with the point spread
function (PSF) fitting program {\it StarFinder} ({\it 18}).  Third, the positions of the 
stars in each epoch were transformed into a common coordinate system.
Fourth, orbital model fits to the positions of stars with significant orbital
phase coverage provided estimates of each star's orbital parameters.

The gravitational potential in this region of the Galaxy is dominated by the black hole, and to a good approximation the stars follow Keplerian orbits. While deviations from a pure Keplerian orbit will likely be detected in the future, this cannot be expected with the current data set, and we therefore assumed a point-mass potential and Newtonian gravity. Within this framework, we used a $\chi^2$-minimization routine to find the best fit orbital parameters ({\it 7}). The six orbital elements describing the star's trajectory are the period, eccentricity, time of closest approach (periapse passage), inclination, angle to periapse, and the angle of the ascending node. The seven parameters describing the gravitational potential are mass, 3-dimensional position (distance from Earth \& focal point of the orbital ellipse) and 3-dimensional velocity of the point-mass.

We have covered a complete orbit of S0-102 with astrometric observations. Therefore, the data contain enough information that they can in principle be used to fit for both the star's orbital elements and also for the parameters of the potential ({\it 19}).  However, this requires the same low level of systematic effects in the data as for S0-2, which dominates our current knowledge about the potential and is a factor of 16 brighter than S0-102.  Considering that fainter sources are much more prone to source confusion than brighter ones, it is unlikely that S0-102 has the same low level of systematic errors ({\it 7, 8}). 

A way of minimizing the effects of possible confusion is to not fit for the potential, but rather fix it to the values that have been derived from the orbit of S0-2 (Table~S4). This left only S0-102's orbital elements free, which were then found by fitting the data (Fig.~2). Overall, the data are well fit by this model, which has a reduced-$\chi^2$ of 2.0. The most discrepant points are the two epochs in 2009, both of which lie $2.3\sigma$ away from the best fit. This is the region around closest approach to the black hole, where Sgr~A* and S0-104 ({\it 20}), which are blended in this epoch, are very close ($\sim$55 mas) to S0-102. This constellation most likely leads to an astrometric bias, which supports our conservative approach of fixing the parameters of the potential. 

We determined uncertainties on S0-102's orbital parameters via Monte Carlo simulations assuming Gaussian error statistics. Using the best fit and the error bars on the measured data points, we generated $10^5$ artificial data sets and fitted an orbit to each realization. The probability distribution functions for the orbital elements are the distribution of best fit values from the Monte Carlo simulations (Table~1). This approach assumes that the positional errors are statistical in nature and does not account for possible systematic contributions like unrecognized source confusion.

Our high angular resolution imaging campaign of the Galactic center started in 1995, constituting a time baseline of 17 years, 7 of which were carried out with deep AO observations. While this is long enough to detect accelerations (i.e. curvature) of several stars within the central arcsecond, it only allows a reliable orbit determination for the shortest period stars. We consider an orbit reliable if it is likely that an increase in data and a decrease in noise does not lead to major changes in the orbital elements.  As a rule of thumb, 50\% of the orbit needs to be covered by observations in order to determine the orbit reliably  ({\it 21}). Our data set traces S0-102 through a complete orbit, making it the only star other than S0-2 for which such a fraction of the orbit is sampled. The key property that enabled this orbital coverage is S0-102's period.

With an orbital period of 11.5 years, S0-102 has the shortest period among all known stars orbiting the black hole in our galaxy's center (Fig.~3). It is 5 years ($\sim$30\%) shorter than S0-2's orbital period, which was the previously known shortest-period star. The period (or, alternatively, the semi-major axis, which is related to the period via Kepler's law) is the most important orbital parameter because it not only makes it possible to sample a substantial fraction of the orbit by observations, but also simplifies the detection of relativistic effects, most of which are cumulative and build up with increasing phase coverage until they breach the detection threshold.

A test of Einstein's General Theory of Relativity is the next goal in Galactic center research now that the existence of a black hole is well established. This theory has so far passed all tests on solar system scales with flying colors. However, the gravitational field of the sun is very weak, with typical experiments probing regimes of up to $\epsilon \sim GM/(Rc^2) \sim 10^{-6}$. Here, $G$ is the gravitational constant, $M$ the mass scale, $R$ the distance scale, and $c$ the speed of light ({\it 22}). The gravitational fields that have been probed in tests using double neutron stars like the Hulse-Taylor binary pulsar are of the same magnitude, since the masses and separation of the neutron stars are comparable to the mass and radius of the Sun ({\it 23}). In the Galactic center, stars like S0-102 and S0-2 probe gravity regimes that are two orders of magnitude stronger, $\epsilon \sim 10^{-4}$.

The effects of curved space time manifest themselves in two ways: the orbit of a star deviates from its Keplerian approximation, and the path and wavelength of the light emitted by it get changed. It has been shown that for S0-2 the relativistic redshift should become measurable at S0-2's next periapse passage in 2018 ({\it 24, 25}). Although this is a measurement of S0-2's redshift, S0-102 will render this measurement more precise. The reason for this is that, with AO observations of S0-102's apoapse passage in the next years, there will be enough information in the data set to lower the level of systematic effects to  an insignificant level, and thereby allow S0-102 to be used to constrain the central gravitational potential. This additional information will then translate into a more precise determination of S0-2's orbital elements and therefore redshift measurement. 

The relativistic redshift will also become observable in S0-102 itself.  For S0-102, the difference between the purely Newtonian and fully relativistic Doppler shifts amounts up to $\sim$90 km/s. This is within the capabilities of upgraded or next-generation near-infrared spectrographs for sources of this magnitude. Since the Keplerian orbital parameters must be known precisely to infer the relativistic contribution to the redshift, only stars with well sampled astrometric orbits can be used to detect deviations from a Newtonian orbit. Additionally, the relativistic redshift is strongest at periapse passage, and S0-2 and S0-102 are the only known stars that pass periapse within the next 10 years and that will have an orbital phase coverage greater than 50\% then. Since S0-102 will pass closest approach 3 years after S0-2, an observation of its relativistic redshift will provide a crucial check on systematic effects in S0-2's result, and test the independence of the gravitational redshift from the interior structure of the star -- a key assumption of General Relativity. 

Other than the gravitational redshift, the leading order General Relativistic effect is the precession of the periapse within the orbital plane ({\it 26--31}). This is a cumulative effect that adds up from orbit to orbit and is therefore much more likely to be detected in stars that complete several orbits during the timescale of an observational campaign. Angelil et al.~({\it 30}) show that the effect of periapse precession scales as $\textnormal{\it period}^{-1}$, quantifying the significance of a short period for the test of General Relativity in an unprecedented regime.

The periapse precession, which leads to rosetta-shaped orbits, consists of two components: a General Relativistic part that leads to a prograde precession, and a Newtonian part that leads to a retrograde precession. The Newtonian deviation of a Keplerian orbit is caused by the inevitable presence of extended dark mass around the black hole, which means that the gravitational potential is not point-like ({\it 27, 31, 32}). This extended mass distribution is thought to mainly consist of stellar remnants -- stellar mass black holes and neutron stars -- supplied by the dense nuclear star cluster ({\it 33, 34}). In order to measure the General Relativistic part only, the detection of the periapse shift in more than one star is required to break the degeneracy.  This makes a star like S0-102 a necessary ingredient for a measurement of the warping of spacetime around a supermassive black hole.   

Given the astrometric precision with which  a star can be located on the detector ($\sim$0.1 mas and $\sim$1 mas for S0-2 and S0-102, respectively) and with which a stable absolute reference frame can be constructed (currently $\sim$0.1 mas/yr), the precession of both S0-2's and S0-102's apoapse -- which amounts to $\sim$1 mas and $\sim$0.3 mas, respectively -- is likely only detectable with the next generation of telescopes like the Thirty Meter Telescope.  In this upcoming era, S0-102 will play as dominating a role as S0-2 plays in today's Keck era.

{\bf References and Notes}
\begin{enumerate}
\item A. Eckart, R. Genzel, {\it Mon. Not. R. Astron. Soc.} {\bf 284}, 576 (1997).
\item A. M. Ghez, B. L. Klein,  M. Morris, E. E. Becklin, {\it Astrophys. J.} {\bf 509}, 678 (1998).
\item A. M. Ghez, M. Morris, E. E. Becklin, A. Tanner, T. Kremenek, {\it Nature} {\bf 407}, 349 (2000).
\item A. Eckart, R. Genzel, T. Ott, T., R. Sch\"odel, {\it Mon. Not. R. Astron. Soc.} {\bf 331}, 917 (2002).
\item R. Sch\"odel {\it et al.}, {\it Nature} {\bf 419}, 694 (2002).
\item A. M. Ghez  {\it et al.}, {\it Astrophys. J.} {\bf 586}, L127 (2003).
\item A. M. Ghez  {\it et al.}, {\it Astrophys. J.} {\bf 689}, 1044 (2008).
\item S. Gillessen  {\it et al.}, {\it Astrophys. J.} {\bf 692}, 1075 (2009).
\item S. Gillessen  {\it et al.}, {\it Astrophys. J.} {\bf 707}, L114 (2009).
\item S0-102 might be identical to the star labeled S55 in ({\it 8}), for which only linear motion is reported there.  
\item K. Matthews, B. T. Soifer, in {\it Astronomy with Infrared
Arrays: The Next Generation}, ed. I. McLean (Kluwer, 1994), Astrophysics and Space Science, v. 190, p. 239
\item K. Matthews, A. M. Ghez, A. J. Weinberger, G. Neugebauer, {\it The Publications of the Astronomical Society of the Pacific} {\bf 108}, 615 (1996).
\item A. M. Ghez  {\it et al.}, {\it Astrophys. J.} {\bf 620}, 744 (2005).
\item J. R. Lu {\it et al.}, {\it Astrophys. J.} {\bf 690}, 1463 (2009).
\item A. M. Ghez  {\it et al.}, {\it Astrophys. J.} {\bf 635}, 1087 (2005).
\item S. D. Hornstein {\it et al.}, {\it Astrophys. J.} {\bf 667}, 900 (2007).
\item A detailed description of the methods is available as supplementary material on {\it Science} Online.
\item E. Diolaiti, {\it et al.}, {\it Astron. Astrophys. Suppl. Ser.} {\bf 147}, 335 (2000).
\item The distance to the black hole and its line-of-sight velocity cannot be determined, because a measurement of the star's radial velocity is needed for this. S0-102 is currently just beyond the spectroscopic limit.
\item S0-104 is a previously unknown detected star; its orbital motion is presented in the supplementary material on {\it Science} Online.
\item W. I. Hartkopf, B. D. Mason, E. Worley, {\it Astron. J.} {\bf 122}, 3471 (2001).
\item C. Will, {\it Living Reviews in Relativity} (available at http://www.livingreviews.org/lrr-2006-3), 2001
\item The strength of the gravitational field we quote here relates to the inter-body potential and therefore orbital velocities. However, we want to note that the strong-field internal gravity of the neutron stars could leave imprints on the orbital motion.
\item S. Zucker, T. Alexander, S. Gillessen, F. Eisenhauer, R. Genzel, {\it Astrophys. J.} {\bf 639}, L21 (2006).
\item R. AngŽlil, P. Saha, {\it Astrophys. J.} {\bf 734}, L19 (2011).
\item M. Jaroszynski, {\it Acta Astronomica} {\bf 48}, 653 (1998).
\item G. F. Rubilar, A. Eckart, {\it Astron. Astrophys.} {\bf 374}, 95 (2001).
\item N. N. Weinberg, M. Milosavljevic, A. M. Ghez, {\it Astrophys. J.} {\bf 622}, 878 (2005).
\item C. M. Will, {\it Astrophys. J.} {\bf 674}, L25 (2008).
\item R. AngŽlil, P. Saha, D. Merritt, {\it Astrophys. J.} {\bf 720}, 1303 (2010).
\item D. Merritt, T. Alexander, S. Mikkola, C. M. Will, {\it Phys. Rev. D} {\bf 81}, 062002 (2010).
\item A. Gualandris, S. Gillessen, D. Merritt, {\it Mon. Not. R. Astron. Soc.} {\bf 409}, 1146 (2010).
\item M. R. Morris, {\it Astrophys. J.} {\bf 408}, 496 (1993).
\item J. Miralda-Escude, A. Gould, {\it Astrophys. J.} {\bf 545}, 847 (2000).

\item S. Yelda {\it et al.}, {\it Astrophys. J.} {\bf 725}, 331 (2010).
\item W. I. Clarkson {\it et al.}, {\it Astrophys. J.} {\bf 751}, 132 (2012).
\item S. D. Hornstein {\it et al.}, {\it Astrophys. J.} {\bf 577}, L9 (2002).
\item S. D. Hornstein, thesis, UCLA (2007).
\item R. Sch\"odel {\it et al.}, (available at http://arxiv.org/abs/1110.2261v3).
\item J. C. Christou, S. T. Ridgway, D. F. Buscher, C. A. Haniff, D. W. McCarthy Jr., in {\it Astrophysics with Infrared Arrays}, ASP Conference Series, Volume 14, 133 (1991).
\item J. Primot, G. Rousset, J. C. Fontanella,  {\it Journal of the Optical Society of America A} {\bf 7}, 1598 (1990).
\item M. G. Petr, V. Coude Du Foresto, S. V. W. Beckwith, A. Richichi, M. J. McCaughrean, {\it Astrophys. J.} {\bf 500}, 825 (1998).

Acknowledgements: We thank the staff of the Keck Observatory for all their help in obtaining the new observations. Support for this work was provided by NSF grant AST 0909218, the Levine-Leichtman Family Foundation, and the W. M. Keck Foundation. RS acknowledges support by the Ram\`on y Cajal programme, by grants AYA2010-17631, AYA2009-13036, and PA1002584 of the Spanish Ministerio de Econom\'ia y Competitividad, and by grant P08-TIC-4075 of the Junta de Andaluc\'ia. The W. M. Keck Observatory is operated as a scientific partnership among the California Institute of Technology, the University of California and the National Aeronautics and Space Administration. The Observatory was made possible by the generous financial support of the W. M. Keck Foundation. The authors wish to recognize and acknowledge the very significant cultural role and reverence that the summit of Mauna Kea has always had within the indigenous Hawaiian community. We are most fortunate to have the opportunity to conduct observations from this mountain. The data described in the paper are presented in the Supporting Online Material.

\end{enumerate}

\clearpage
\pagebreak

\begin{deluxetable}{lc}
\tabletypesize{\scriptsize}
\tablewidth{0pt}
\tablecaption{Orbital elements for S0-102 \tablenotemark{a} \label{tbl1}}
\tablehead{
        \colhead{Parameter [Unit]} &
	\colhead {Value} 
}
\startdata
\sidehead{S0-102's orbital parameters }  %%\tablenotemark{a}
\hline
\noalign{\vskip .5ex}
Period [years] & $11.5 \pm 0.3$ \\
Time of Closest Approach [year] & $2009.5 \pm 0.3 $ \\
Eccentricity & $0.68 \pm 0.02$\\
Inclination\tablenotemark{b,c} \,\,\,\, [degrees] & $151 \pm 3$ \\
Angle to Periapse [degrees] & $185\pm 9$ \\
Position Angle of the Ascending Node\tablenotemark{c} \, [degrees] & $175 \pm 5$ \\
\cutinhead{Parameters of the potential\tablenotemark{d} }
Mass [$10^6$ M$_{\textnormal {Sun}}$] & $4.1 \pm 0.4$ \\
Distance [kpc] & $7.7 \pm 0.4$ \\
 \enddata
\tablenotetext{a}{The best fit has a $\chi^2$ of 39.96 with 20 degrees-of-freedom.}
\tablenotetext{b}{90 deg is edge-on, and 0 deg is face-on.}
\tablenotetext{c}{The allowed ranges for inclination and angle of the ascending node are [0,180].}
\tablenotetext{d}{The parameters that describe the gravitational potential have been taken from S0-2's orbit. We list here mass and distance, see Table~S4 for 2D position and 3D velocity of the central mass.}
\end{deluxetable}

\clearpage

\begin{figure}
\includegraphics[width=10.cm]{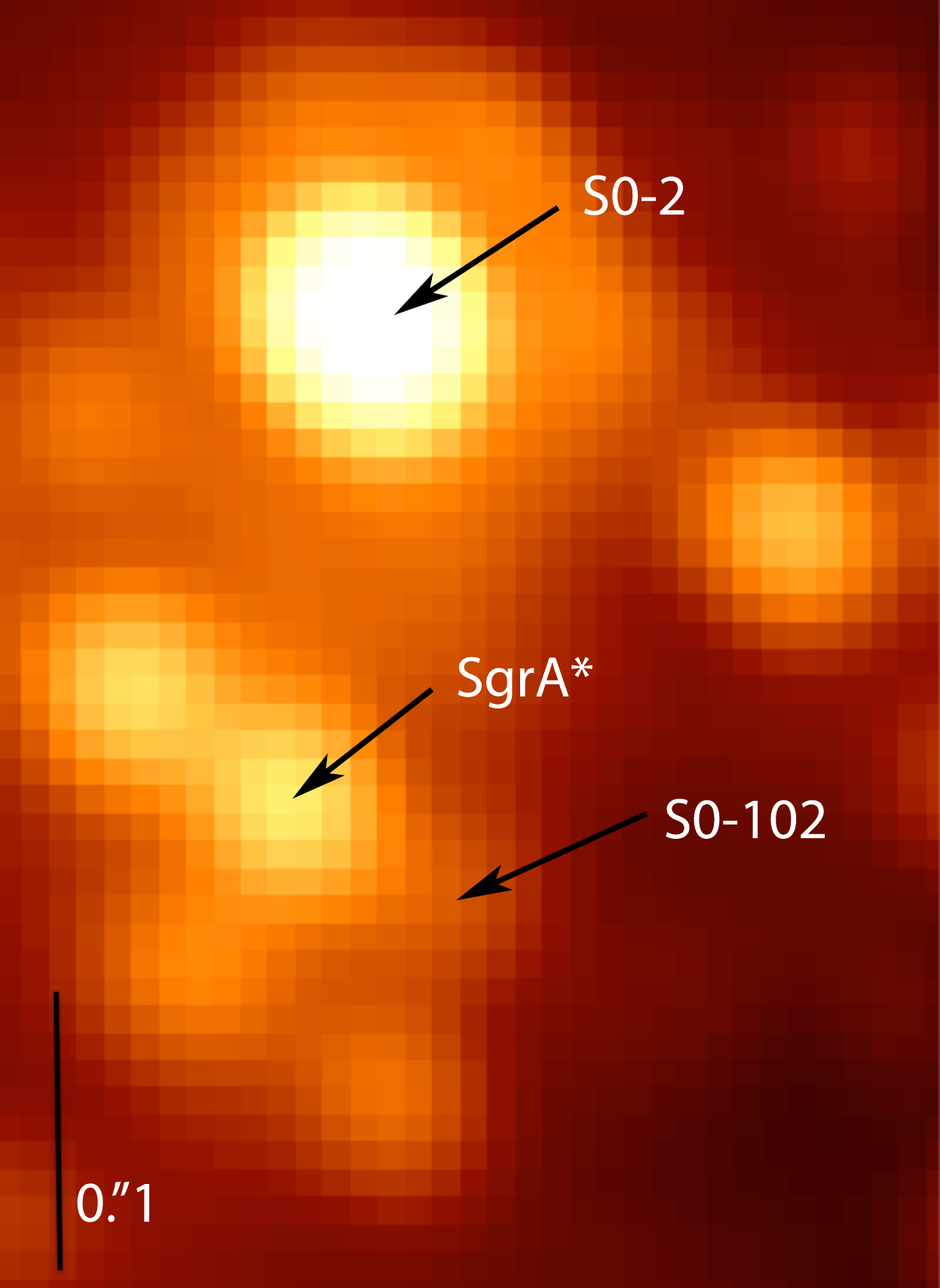}
\caption{
%{\bf Fig. 1:} 
A Keck/NIRC2 adaptive optics image from May 2010 showing the short-period star S0-102, which is besides S0-2 the only star with full orbital phase coverage, and the electromagnetic counterpart of the black hole, Sgr~A*. The image was taken at a wavelength of 2.12 $\mu$m and shows the challenge of detecting S0-102, which is 16 times fainter than S0-2 and lies in this crowded region.}
\end{figure}

\begin{figure}
\includegraphics[width=15.5cm]{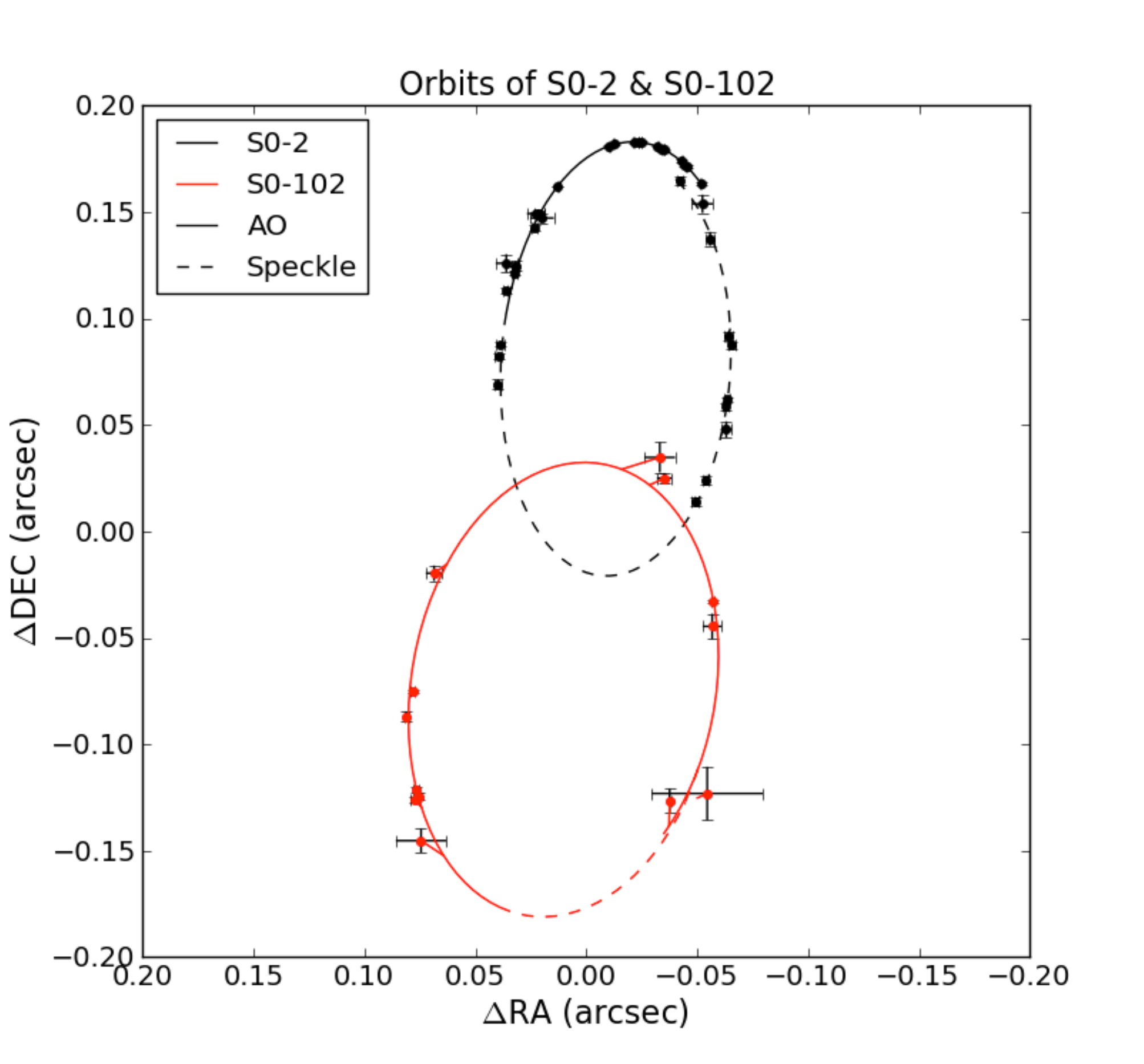}
\caption{
%{\bf Fig. 2:} 
The orbits of S0-2 (black) and S0-102 (red). The data points and the best fits are shown. Both stars orbit clockwise. The dashed lines represent the part of the orbits that have been observed with Speckle data, the solid lines indicate adaptive optics observations. The data points for S0-2 range from the year 1995 to 2012, S0-102's detections range from 2000 to 2012. The connecting lines to the best fit visualize the residuals. Note that while the best-fit orbits are not closing, the statistically allowed sets of orbital trajectories are consistent with a closed orbit. S0-102 has an orbital period of 11.5 years, 30\% shorter than S0-2, the previously known shortest-period star.}
\end{figure}

\begin{figure}
\includegraphics[width=15.cm]{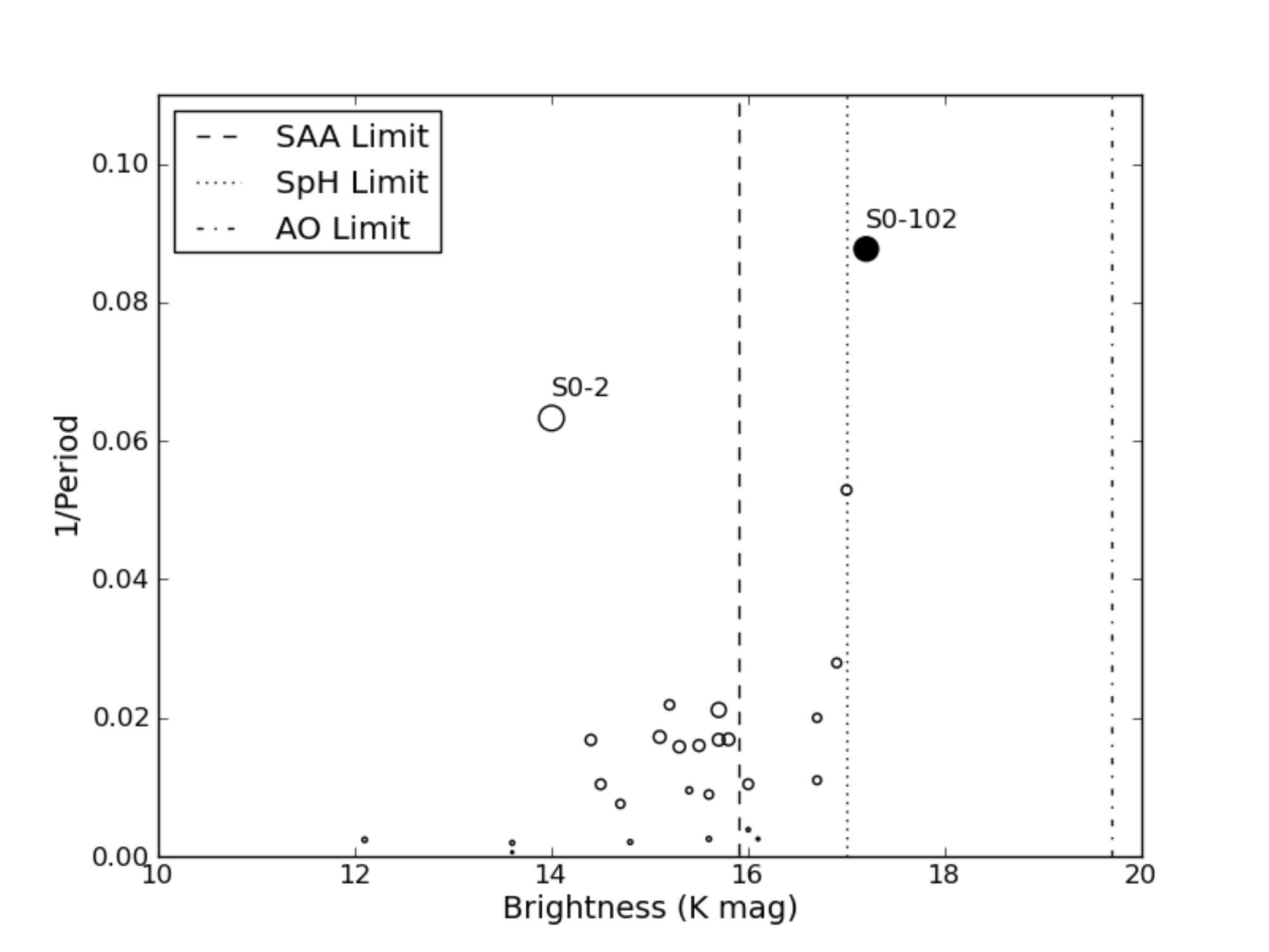}
\caption{
%{\bf Fig. 3:}
 The orbital periods and magnitudes of all known stars orbiting the Galactic black hole ({\it 7, 8,} this publication). The size of the points is scaled with the orbital phase that is covered by observations. The higher the fraction of the orbit that is sampled by observations the more reliable the orbital solution is. The three vertical lines indicate, from left to right, the limiting magnitude for Speckle shift-and-add (SAA), Speckle holography (SpH), and adaptive optics (AO) data sets. We define the limiting magnitude as the median of the limiting magnitudes of individual epochs, which are listed in Tables~S2 \& S3. The filled point represents the star S0-102.  }
\end{figure}

\end{document}